\documentclass[a4paper,fleqn]{cas-dc}

\usepackage[numbers, sort&compress]{natbib}

\usepackage{algorithm}  
\usepackage{algorithmic}  

\usepackage{amsmath} 
\usepackage{bm}      
\newtheorem{theorem}{Theorem}
\usepackage{caption}
\begin{document}
\let\WriteBookmarks\relax
\def\floatpagepagefraction{1}
\def\textpagefraction{.001}


\shortauthors{Xumin Pu, Zhinan Sun et al.}  

\title[mode = title]{Low-Complex Channel Estimation in Extra-Large Scale MIMO with the Spherical Wave Properties}




\author[1,3]{Xumin Pu}
  \ead{puxm@cqupt.edu.cn}
\credit{Conceptualization, Funding acquisition, Supervision, Writing}

\author[2]{Zhinan Sun}
\ead{szn@my.swjtu.edu.cn}
\credit{Investigation, Software, Writing}

\author[4]{Qianbin Chen}\ead{chenqb@cqupt.edu.cn}
\credit{Funding acquisition, Project administration}

\author[1]{Shi Jin}\cormark[1] \ead{jinshi@seu.edu.cn}
\credit{Funding acquisition, Project administration}

\address[1]{National Mobile Communications Research Laboratory, Southeast University, Nanjing 210096, China}
\address[2]{School of Information Science and Technology, Southwest Jiaotong University, Chengdu 611756, China}
\address[3]{School of Communications and Information Engineering, Chongqing University of Posts and Telecommunications, Chongqing 400065, China}
\address[4]{Chongqing Key Laboratory of Mobile Communications Technology, Chongqing 400065, China}

\cortext[cor1]{Corresponding author}

\begin{abstract}
	This paper investigates the low-complex linear minimum mean squared error (LMMSE) channel estimation in an extra-large scale MIMO system with the spherical wave model (SWM). We model the extra-large scale MIMO channels using the SWM in the terahertz (THz) line-of-sight propagation, in which the transceiver is a uniform circular antenna array. On this basis, for the known channel covariance matrix (CCM), a low-complex LMMSE channel estimation algorithm is proposed by exploiting the spherical wave properties (SWP). Meanwhile, for the unknown CCM, a similar low-complex LMMSE channel estimation algorithm is also proposed. Both theoretical and simulation results show that the proposed algorithm has lower complexity without reducing the accuracy of channel estimation.
\end{abstract}


\maketitle
\section{Introduction}\label{sec1}

Due to substantial improvement in spectral efficiency and energy efficiency, extra-large scale multiple input multiple output (MIMO) systems have attracted much attention for the sixth-generation (6G) wireless networks.  When the array size becomes such a large
dimension, the users and scatterers could be located inside the Rayleigh distance of the large arrays. Under this condition, the spherical wavefronts are experienced over the arrays. In addition, due to the demand for spectrum resources, the 6G communication focuses on the terahertz (THz) band. For the accurate modeling of the THz MIMO channels, the spherical wavefronts must be considered \cite{Noh}.

Some work with the spherical wave model (SWM) has been presented~\cite{pu,Wu,Pu8,Ai,Magoarou,Han7}. Considering
the joint effects of path loss and phase differences with the SWM, the optimal antenna placement in the line of sight (LoS) channels was investigated~\cite{pu}. For the massive MIMO channels, the authors proposed a general channel model with the spherical wavefront~\cite{Wu}. Meanwhile, the transmit design with the SWM was analytically investigated in~\cite{Pu8}. In \cite{Ai}, a scatterer localization algorithm with the SWM was proposed. In addition, the researchers in~\cite{Magoarou} studied the channel estimation algorithms with the SWM in a multi-user MIMO scenario. Recently, for the extra-large scale MIMO systems with the SWM, \cite{Han7} proposed two channel estimators based on subarray-wise and scatterer-wise methods, in which the multipath channel with the last-hop scatterers is modelled as the SWM.

However, these previous studies did not consider the low-complex transceiver design by utilizing the spherical wave properties (SWP) in the extra-large scale MIMO systems. As a particularly important part of the transceiver design, the low-complex channel estimator is the focus for this paper. As an excellent estimator, linear minimum mean squared error (LMMSE) effectively employs the statistical information of the wireless channel to achieve the optimal mean square error \cite{Savaus}. The LMMSE channel estimator has achieved good performance for any signal-to-noise ratio (SNR) condition \cite{Jacobs2009}. In addition, since the LMMSE estimator is orthogonal to its estimation error, the analysis of the capacity bounds becomes simple \cite{Taesang2004}. Recently, there is still a lot of research on the LMMSE channel estimator \cite{wu2021, Wan, Cheng}. A vector quantization method for the LMMSE channel estimation was proposed in \cite{wu2021}, which calculates the LMMSE filter matrix of some typical wireless channels off-line. In order to reduce the hardware cost and the power consumption, \cite{Wan} proposed a bussgang LMMSE channel estimator for the one-bit quantization massive MIMO systems. For some advanced Bayesian channel estimation algorithms \cite{Cheng}, the LMMSE estimator is also the basis of the algorithm iteration steps. These papers show that the research of the LMMSE channel estimator is still meaningful today. Meanwhile, for the extra-large scale MIMO systems, the LMMSE channel estimator will not be simple to implement in practical systems because of the high computational overhead for the high-dimensional matrix inversion. How to obtain the compromise between the performance and the complexity of the LMMSE channel estimator has become the practical challenge in extra-large scale MIMO systems. 

There have been some studies on low-complex LMMSE channel estimators in the orthogonal frequency division multiplexing (OFDM) systems \cite{Edfors, Zhou}. In \cite{Edfors}, a low-rank estimator for the OFDM systems was proposed by using the singular value decomposition (SVD), which exploited the theory of optimal rank-reduction. Based on the FFT of the channel in the delay domain, a low complexity LMMSE channel estimator using the circulant structure of the channel covariance matrix (CCM) was proposed in \cite{Zhou}.  However, these schemes can not adapt to extra-large scale MIMO systems, especially the spherical wavefronts experienced over the arrays. Motivated by the low-complex MMSE estimator in \cite{r11}, this paper presents a low-complex LMMSE channel estimator for extra-large scale MIMO systems by exploiting the channel matrix circulant structure with the spherical wavefronts.

In the 6G wireless communication systems, the demand for the higher data rates will lead to the allocation of wider bandwidth in the THz frequency range. In the THz band, the roughness of most surfaces (e.g. concrete walls) is comparable to  the wavelength, so the multipath component in the THz band is weak \cite{JSong}. As a result, THz communication is mainly focused on the LoS propagation, which makes it suitable for many emerging scenarios, such as the wireless backhaul networks \cite{Do}. The uniform circular array (UCA) can be effectively applied to this communication scenario. On the one hand, the UCA has been widely studied in the wireless communication systems, especially in the LoS MIMO systems \cite{PWang}. On the other hand, by fixing the transmitter and receiver at two locations, the UCA-based transceiver is considered as a candidate for the wireless backhaul communications \cite{Jing,Long}.  Motivated by these observations, in this paper we study the low-complex LMMSE channel estimator by exploiting the circulant structure introduced by the combination of the SWM and UCA.

Our contributions are summarized as follows:
\begin{itemize}
	\item In this paper, for the wireless backhaul networks in the THz LoS channel, we propose a low-complex extra-large scale MIMO channel estimator based on the UCA transceiver. As far as the authors know, this paper is the first to consider the low-complex channel estimator in this scenario, which has practical significance for the 6G wireless communication.
	
	\item In addition, this paper has the unique contributions to the proposed scenario, which are different from some previous studies in the orthogonal frequency division multiplexing (OFDM) systems.  Compared with \cite{Edfors}, our low-complex LMMSE channel estimator exploits the circulant structure introduced by the SWM, which makes the discrete Fourier transform (DFT) matrix can be used for the eigenvalue decomposition of the CCM, thus the complexity of the matrix inversion is significantly reduced by the fast Fourier transform (FFT). 
	Compared with \cite{Zhou}, our contributions are still quite different.  Firstly, the circulant structure in our paper comes from the ingenious combination of the SWM and the UCA. Specifically, this circulant structure is observed in the spatial domain instead of the circulant CCM in the frequency domain and delay domain in \cite{Zhou}.  Secondly, the low-complex channel estimator proposed in our paper is suitable for the extra-large scale MIMO systems, which can not be achieved in \cite{Zhou}.  Since the dimension of the circulant CCM in our paper is related to the size of the antenna array, the complexity of the proposed LMMSE estimator can be reduced to the order of $\mathcal{O}\left(N \log _{2} N\right)$, where $N$ is the number of the antennas. 
	Obviously, our proposed scheme has a significant advantage of low complexity in the extra-large scale MIMO systems. 
	
	\item Further, the low-complex LMMSE channel estimators for both known CCM and unknown CCM are given. For the known CCM, benefiting from the SWP, the computational complexity in terms of  multiply and add  operations (MADs) can be reduced from $3{{N}^{3}}+N$ to $3N\log_2 N+2N$. When the number of antennas $N=512$, our proposed algorithm can reduce the computational complexity by thousands of times. For the unknown CCM, benefiting from the SWP, the computational complexity in terms of MADs can be reduced from $(2 T+3) N^{3}-T N^{2}+N$ to  $2 T N^{2}+3 N \log _{2} N+(3-T) N+T-1$. When the number of antennas $N=512$ and the number of slots $T=10$, our proposed algorithm can reduce the computational complexity by hundreds of times. Therefore, the channel estimation scheme proposed in this paper is helpful for the development of the extra-large scale MIMO from the theory to the practical application.
	
\end{itemize}

\emph{Notations}: Throughout this paper, lowercase and uppercase bold letters represent vectors and matrices, respectively. The operation $(\cdot)^H$ and $(\cdot)^{-1}$ denote the conjugate transpose and matrix inversion, respectively. $\mathbb{E}[\cdot]$ denotes the expectation and $\mathcal{C N}\left(\bm{x}, \bm{R}\right)$ denotes the complex Gaussian function with mean $\bm{x}$ and covariance $\bm{R}$.

\emph{Outline}: The remainder of this paper is as follows: Section 2 introduces the system model of the communication scenario. In Section 3, the low-complex LMMSE channel estimation algorithm based on spherical wave model is introduced in detail. In Section 4, the algorithms are verified by simulations. Finally, Section 5 summarizes the work of the whole paper.
\section{System Model}

\begin{figure*}[tbp]
	\centering
	\includegraphics[scale=0.8]{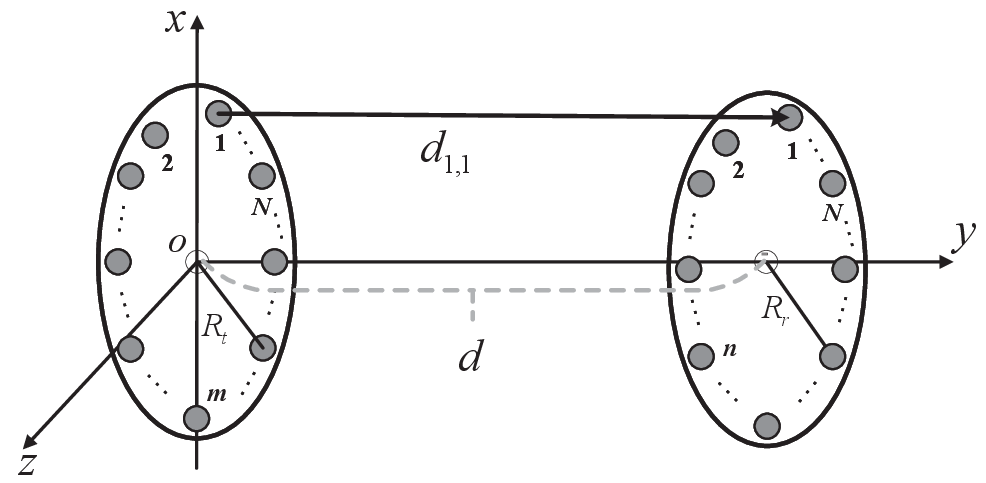}
	\caption{An extra-large scale MIMO system with a $N$-element uniform circular array (UCA) both at the transmitter and at the receiver}
	\label{fig.1.}
\end{figure*}

Consider an extra-large scale MIMO system with a $N$-element UCA both at the transmitter and at the receiver which are parallel to each other, as shown in the Fig.~\ref{fig.1.}. The transmitter with the UCA and its center ${O}$ can be assumed to be located in the
$xz$-plane and at the origin, respectively. The receiver with the UCA is parallel to the transmitter and its center is coaxial with the $y$ axis. ${R_t}$ and ${R_r}$ are the radius of the UCA at the transmitter and at the receiver, respectively. Due to the lack of diffraction in the THz band, the radio propagation mainly focuses on the LoS path, which is the basis of the channel model in this paper.
By taking the SWM \cite{Magoarou,Torkildson} into account, the elements of channel matrix for  extra-large scale MIMO systems can be written as
\begin{equation}\label{channel response1}
[\bm{H}]_{n, m}=\frac{\lambda}{4 \pi d_{n, m}} e^{j \frac{2 \pi}{\lambda} d_{n, m}},
\end{equation}
where $\lambda$ is the wavelength, $d_{{n,m}}$ denotes the distance between the $m$th transmit antenna $\left(m=1,\dots,N\right)$ and the $n$th receive antenna $\left(n=1,\dots,N\right)$.

By applying the similar method in  \cite{Pu8}, $d_{{n,m}}$ can be derived by
\begin{equation}\label{distance}
\begin{aligned}
d_{{n,m}}\approx {{d}}+\frac{R_{t}^{2}+R_{r}^{2}-2{{R}_{t}}{{R}_{r}}\cos (\frac{2\text{ }\!\!\pi\!\!\text{ }}{N}(n-m))}{2{{d}}},
\end{aligned}
\end{equation}
where $d$ is the distance between the two parallel arrays. Substituting $d_{{n,m}}$ into \eqref{channel response1}, the channel response between the $n$th receiving antenna and the  $m$th transmitting antenna is obtained as
\begin{equation}\label{channel response2}
\begin{aligned}
{[\bm{H}]_{n, m} } \!\!=\!\alpha_{n, m} \cdot 
\exp  (j \pi \frac{2 d^2\!\!+\!\!R_t^2\!\!+\!\!R_r^2\!\!-\!\!2 R_t R_r \cos\! \left(\frac{2 \pi}{N}(n\!\!-\!\!m)\!\right )}{\lambda d}\!),
\end{aligned}
\end{equation}
where $\alpha_{n, m}=\frac{\lambda d}{4 \pi d^2+2 \pi\left(R_t^2+R_r^2-2 R_t R_r \cos \left(\frac{2 \pi}{N}(n-m)\right)\right)}$.

In the training phase for channel estimation, the user sends
$L$-length orthogonal pilot sequences. We assume that the receiver obtains $T$ independent observations in each coherence interval. The pilot
sequences transmitted by the $N$ antennas in the $t$th slot $\left(t=1,\dots,T\right)$ can be denoted by a $L \times N$ matrix ${{\boldsymbol{\Gamma }}_t}$ with $\boldsymbol{\Gamma}_{t}^{H} \boldsymbol{\Gamma}_{t}=\boldsymbol{I}_{N}$. The base station receives the $N\times L$ signal in the $t$th slot as
\begin{equation}\label{the estimate channel}
\bm{Y}_{t}=\bm{H}_{t} \boldsymbol{\Gamma}_{t}^{H}+\bm{N}_{t}, \quad t=1, \ldots, T
\end{equation}
where $\bm{N}_{t}$ is the $N\times L$ additive white Gaussian noise (AWGN) matrix. Given ${\bm{Y}_{t}}$ and $\boldsymbol{\Gamma}_{t}$, the goal of channel estimation is to recover $\bm{H}_{t}$. After correlating the received signals with the pilot sequences, we get the observations
\begin{equation}\label{signal}
\tilde{\bm{Y}}_{t}=\bm{H}_{t}+\tilde{\bm{N}}_{t}, \quad t=1, \ldots, T
\end{equation}
where $\tilde{\bm{Y}}_{t}={\bm{Y}_{t}}{{\boldsymbol{\Gamma }}_{t}}$, and $\tilde{\bm{N}}_{t}={\bm{N}_{t}}{{\boldsymbol{\Gamma }}_{t}}$ is a noise matrix with independent and identically distributed (i.i.d.) zero-mean and element-wise variance ${\sigma }^{2}$. We have  $\bm{H}_{t} \sim \mathcal{C} \mathcal{N}\left(\mathbf{0}, \bm{R}_{\bm{H}}\right)$ with $\bm{R}_{\bm{H}}=\mathbb{E}\left[\bm{H}_{t}^{H} \bm{H}_{t}\right]$ and $\tilde{\bm{N}}_{t} \sim \mathcal{C} \mathcal{N}\left(\mathbf{0}, \bm{R}_{\tilde{\bm{N}}}\right)$ with
$\bm{R}_{\tilde{\bm{N}}}=\sigma^{2} \bm{I}_{\emph{N}}$.

\section{Low-complex LMMSE Channel Estimation with the SWM}
In this section, by exploiting the circulant structure
introduced by the SWM, we propose a low-complex LMMSE channel estimation.

By considering a LMMSE estimator in matrix form that is derived from \cite{M}, the channel matrix $\bm{H}_{t}$  in (\ref{signal}) can be expressed as

\begin{equation}\label{channel estimation}
\begin{aligned}
{\hat{\bm{H}}_{t}^{\text {LMMSE }}}=\tilde{\bm{Y}}_{t}\bm{A},
\end{aligned}
\end{equation}
where
\begin{equation}\label{LMMSE matrix}
\begin{aligned}
\bm{A}={{\text{(}{{\bm{R}}_{\bm{H}}}+{\bm{R}_{\tilde{\bm{N}}}\text{)}}^{\text{-}1}}{{\bm{R}}_{\bm{H}}}}.
\end{aligned}
\end{equation}
The computational complexity of the matrix ${{\text{(}{{\bm{R}}_{\bm{H}}}+{\bm{R}_{\tilde{\bm{N}}}}\text{)}}^{\text{-}1}}$ in (\ref{LMMSE matrix}) is $\mathcal{O}\left(N^{3}\right)$. For an extra-large scale MIMO system, the computation complexity is excessively high due to the large number of antennas. In addition, since the time-varying characteristics of the channel, this matrix inversion operation is frequently updated. Hence, for the extra-large scale MIMO system, the low-complex LMMSE channel estimation needs to be  investigated.

In the investigation that follows, for the known channel covariance matrix, a low-complex LMMSE channel estimation algorithm is proposed by exploiting the circulant structure introduced by the SWM.
Meanwhile, for the unknown channel covariance matrix, a similar low-complex LMMSE channel estimation algorithm is also proposed.

\subsection{Known Channel Covariance Matrix}
Here we denote the channel covariance matrix with the SWM by ${{\bm{R}}_{\bm{H}}}$. And the complex conjugate operation is denoted by $\overline {( \cdot )} $. Furthermore, the entry of ${{\bm{R}}_{\bm{H}}}$ in the $m$th row $\left(m=1,\dots,N\right)$ and the $n$th column $\left(n=1,\dots,N\right)$  can be derived by
\begin{equation}\label{the entry}
\begin{aligned}
\left[{{\bm{R}}_{\bm{H}}}\right]_{m, n} &=\mathbb{E}\left(\sum_{i=1}^{N} \overline{\left([\bm{H}]_{i, m}\right)}[\bm{H}]_{i, n}\right) \\
&=\sum_{i=1}^{N}  {\alpha_{i, m} \alpha_{i, n}} \exp \left(\frac{j 2 \pi\left(d_{i, n}-d_{i, m}\right)}{\lambda}\right)\\
&=\sum_{i=1}^{N}  { w_{m,n}^i} \exp \left(\frac{j 2 \pi\left(d_{i, n}-d_{i, m}\right)}{\lambda}\right),
\end{aligned}
\end{equation}
where $w_{m, n}^i=\alpha_{i, m} \cdot \alpha_{i, n}$ and
\begin{equation}
d_{i,m}-d_{i,n}=\frac{{{R}_{r}}{{R}_{t}}\left[ \cos \frac{2\pi (i-n)}{N} \right.\left. -\cos \frac{2\pi (i-m)}{N} \right]}{{d}}.
\end{equation}
From (\ref{the entry}), ${{\bm{R}}_{\bm{H}}}$ can be written as (\ref{the whole covariance matrix}), which is shown at the next page. Furthermore, it is shown that the matrix ${{\bm{R}}_{\bm{H}}}$ is a circulant matrix with ${{[{{\bm{R}}_{\bm{H}}}]}_{m,n}}={{[{{\bm{R}}_{\bm{H}}}]}_{m+1,n+1}}$.

\newcounter{TempEqCnt}
\setcounter{TempEqCnt}{\value{equation}}
\setcounter{equation}{9}
\begin{figure*}[   b]
	
	\begin{equation}\label{the whole covariance matrix}
	{{\boldsymbol{R}}_{\boldsymbol{H}}}\!\!=\!\!\!\left[ \begin{aligned}
	& \sum\limits_{i=1}^{N}{{ w_{1,1}^i \!\exp \left( \frac{j2\pi }{\lambda }(d_{i,1}-d_{i,1}) \right)}}\ \  \sum\limits_{i=1}^{N}{{ w_{1,2}^i \!\exp \left( \frac{j2\pi }{\lambda }(d_{i,1}-d_{i,2}) \right)}} \cdots \sum\limits_{i=1}^{N}{{ w_{1,N}^i \!\exp \left( \frac{j2\pi }{\lambda }(d_{i,1}-d_{i,N}\!) \right)}}\  \\
	& \sum\limits_{i=1}^{N}{{ w_{2,1}^i \!\exp  \left( \frac{j2\pi }{\lambda }(d_{i,2}-d_{i,1}) \right)}}\ \ \sum\limits_{i=1}^{N}{{ w_{2,2}^i \!\exp  \left( \frac{j2\pi }{\lambda }(d_{i,2}-d_{i,2}) \right)}} \cdots \sum\limits_{i=1}^{N}{{ w_{2,N}^i \!\exp  \left( \frac{j2\pi }{\lambda }(d_{i,2}-d_{i,N}\!) \right)}} \\
	& \ \ \ \ \ \ \ \ \ \ \ \ \ \ \vdots \ \ \ \ \ \ \ \ \ \ \ \ \ \ \ \ \ \ \ \ \ \ \ \ \ \ \ \ \ \ \ \ \ \ \ \  \vdots  \ \ \ \ \ \ \ \ \ \ \ \ \ \ \ \ \ \ \ \ \ \ \ \ \ \ \ \ \ \ \ \ddots  \ \ \ \ \ \ \ \ \ \ \ \ \ \ \ \ \ \ \ \ \ \ \ \ \ \ \ \ \vdots  \\
	& \sum\limits_{i=1}^{N}\!\!{{ w_{N,1}^i \!\exp  \left( \frac{j2\pi }{\lambda }(d_{i,N}-d_{i,1}) \right)}}\ \sum\limits_{i=1}^{N}\!\!{{ w_{N,2}^i \!\exp  \left( \frac{j2\pi }{\lambda }(d_{i,N}-d_{i,2}) \right)}} \cdots \sum\limits_{i=1}^{N}\!\!{{ w_{N,N}^i \!\exp  \left( \frac{j2\pi }{\lambda }(d_{i,N}-d_{i,N}\!)\! \right)}} \\
	\end{aligned}\!\! \right]
	\end{equation}
\end{figure*}
\setlength{\parskip}{0\baselineskip}

\subsubsection{Spherical wave properties are utilized}
Based on this, the low-complex LMMSE channel estimation based on SWP  is provided in the following theorem.
\begin{theorem}
	For the extra-large scale MIMO system, using the circulant structure introduced by the SWM, the LMMSE estimate of the channel matrix $\bm{H}_{t}$ in (\ref{signal}) could be calculated as
	\begin{equation}\label{the LMMSE estimate}
	\hat{\bm{H}}_{t}^{\text {LMMSE }}=\tilde{\bm{Y}}_{t} \bm{F}_{{\emph{N}}}\left(\boldsymbol{\Omega}+\sigma^{2} \bm{I}_{N}\right)^{-1} \boldsymbol{\Omega} \bm{F}_{{\emph{N}}}^{H} ,
	\end{equation}
	where ${{\bm{F}}\!_{{\emph{N}}}}\in {\mathbb C}^{N\times N}$ is a discrete Fourier transform (DFT) matrix, ${\boldsymbol{\Omega}}{\rm{ = diag}}\left\{ {{r_1},{r_2}, \ldots ,{r_{{N}}}} \right\}$ is the eigenvalue matrix of ${{\bm{R}}_{\bm{H}}}$ and ${r_k}$ is denoted by
	\begin{equation}\label{r_k}
	r_{_k}=\sum\limits_{n=1}^{N}{{\left[{{\bm{R}}_{\bm{H}}}\right]}_{1,n}}{\rm exp}\left({\frac{-j2\pi nk}{N}}\right)\!\!,\ k=1,2,\dots,N.
	\end{equation}
	\label{thm-1}
\end{theorem}
\setlength{\parskip}{0\baselineskip}

\emph{Proof:}
Since the DFT matrix ${{\bm{F}}\!_{\emph{N}}}$ can be used as the eigenvectors of the circulant matrices \cite{{r10}}, ${{\bm{R}}_{\textbf{H}}}$ can be expressed as
\begin{equation}\label{covariance matrix A}
\begin{aligned}
{{\bm{R}}_{\bm{H}}}=\bm{F}_{\emph{N}}{\boldsymbol{\Omega}}{{\bm{F}}_{{\emph{N}}}^{H}},
\end{aligned}
\end{equation}
where ${{\bm{F}}\!_{\emph{N}}}\in {\mathbb C}^{N\times N}$ is a discrete Fourier transform (DFT) matrix and ${\boldsymbol{\Omega}}{\rm{ = diag}}\left\{ {{r_1},{r_2}, \ldots ,{r _{N}}} \right\}$ is the eigenvalue matrix. And these eigenvalues can be obtained by discrete Fourier transform of the first row of ${{\boldsymbol{R}}_{\boldsymbol{H}}}$\cite{r10}. Thus ${r_{k}}$ is then derived by (\ref{r_k}).

Furthermore, according to (\ref{covariance matrix A}) and after derivations, ${{\text{(}{{\boldsymbol{R}}_{\boldsymbol{H}}}+{{\sigma }^{2}}{{\boldsymbol{I}}_{\emph{N}}}\text{)}}^{\text{-}1}}{{\boldsymbol{R}}_{\boldsymbol{H}}}$ in (\ref{LMMSE matrix}) can be expressed as
\begin{equation}\label{covariance matrix B}
{{\text{(}{{\boldsymbol{R}}_{\boldsymbol{H}}}+{{\sigma }^{2}}{{\boldsymbol{I}}_{{\emph{N}}}}\text{)}}^{\text{-}1}}{{\boldsymbol{R}}_{\boldsymbol{H}}}=\boldsymbol{F}_{\emph{N}}{{\left({\boldsymbol{\Omega}}+{{\sigma }^{2}}\boldsymbol{I}_{\emph{N}}\right)}^{-1}}{\boldsymbol{\Omega}}{{\boldsymbol{F}}_{\emph{N}}^{H}}.
\end{equation}
Substituting (\ref{covariance matrix B}) into (\ref{channel estimation}), we can get \emph{Theorem 1}.

\subsubsection{Spherical wave properties are not utilized}
Considering the comparison of performance and complexity, we also analyze the derivation without circulant structure. In this case, the channel covariance is known, so we directly use cholesky decomposition to complete the matrix inversion in (\ref{LMMSE matrix}). This case is actually the original LMMSE scheme, which will not be repeated here.

From \emph{Theorem 1}, the proposed method circumvents the matrix inversion by exploiting the circulant structure  introduced by the SWM.  Compared with the case without circulant structure, the computational complexity will be reduced from $3{{N}^{3}}+N$ to $3N\log_2 N+2N$. The reduction on  the computation complexity is very meaningful.
For example, when $N=256$, we have $\left({3{{N}^{3}}+N}\right)\big/ \left({3N\log_2 N+2N}\right)=7561.9$. This implies that the computational complexity is reduced by more than 7560 times. Therefore, the proposed scheme has great computational advantage over the conventional methods in extra-large scale MIMO system.
\subsection{Unknown Channel Covariance Matrix}
In practical scenarios, the channel covariance matrix ${{\boldsymbol{R}}_{\boldsymbol{H}}}$ is not necessarily known. In this case, ${{\boldsymbol{R}}_{\boldsymbol{H}}}$ needs to be estimated in advance. We still derive it in two cases according to whether the matrix circulant structure introduced by the SWM is utilized or not.
\subsubsection{Spherical wave properties are utilized}
In this case, the circulant structure is still utilized, which means the matrix inversion will be simplified. Inspired by \eqref{the LMMSE estimate},  ${{\boldsymbol{R}}_{\boldsymbol{H}}}$ is a circulant matrix, so we only need to estimate the eigenvalues of ${{\boldsymbol{R}}_{\boldsymbol{H}}}$ rather than itself.

Exploiting the spherical wave properties, when the CCM is unknown, the LMMSE channel estimation is provided in the following theorem.
\begin{theorem}
	For the extra-large scale MIMO system, using the circulant structure introduced by the SWM, the LMMSE estimate of the channel matrix $\textbf{H}_{t}$ with the unknown CCM could be calculated as
	\begin{equation}\label{the LMMSE estimate WITH}
	\begin{aligned}
	\hat{\boldsymbol{H}}_{t}^{\text {LMMSE }}=\tilde{\boldsymbol{Y}}_{t}\boldsymbol{F}_{\emph{N}}({{{\boldsymbol{\Lambda}^{*}}})^{-1}}{{\left({\boldsymbol{\Lambda}^{*}}-{{\sigma }^{2}}\boldsymbol{I}_{\emph{N}}\right)}}{{\boldsymbol{F}}_{\emph{N}}^{H}},
	\end{aligned}
	\end{equation}
	where ${\boldsymbol{\Lambda}^{*}}{\rm{ = diag}}\left\{ {{\lambda^{*}_{_1}},\ldots{\lambda^{*}_{_i}}, \ldots ,{\lambda^{*}_{N}}} \right\}$
	and $\lambda^{*}_{i}$ is denoted by
	\begin{equation}
	\lambda^{*}_{i}=\sum_{n=1}^{N}\left[\hat{\boldsymbol{R}}_{\tilde{\boldsymbol{Y}}}\right]_{1, n} \exp \left(\frac{-j 2 \pi n k}{N}\right),
	\end{equation}
	\label{thm-1}
	where $ \hat{\boldsymbol{R}}_{\tilde{\boldsymbol{Y}}}=\frac{1}{T} \sum_{t=1}^{T} \tilde{\boldsymbol{Y}}_{t} \tilde{\textbf{Y}}_{t}^{H} $ is the sample covariance matrix.
\end{theorem}

\emph{Proof:}
In the case of unknown CCM, an effective method is to use the maximum likelihood (ML) estimate of the CCM to obtain the LMMSE estimator \cite{r11}. However, if the circulant structure introduced by SWM is considered, we could use the ML estimate of the eigenvalues for CCM rather than itself.

Specifically, a likelihood function for mutually independent observations based on \eqref{signal} is expressed as
\begin{equation}\label{likelihood function}
\begin{aligned}
&L\left(\boldsymbol{\Psi}=\tilde{\boldsymbol{Y}}_{1}, \ldots \tilde{\boldsymbol{Y}}_{t}, \ldots, \tilde{\boldsymbol{Y}}_{T} \mid \boldsymbol{R}_{\boldsymbol{H}}\right) \\
&=\frac{1}{\pi^{T N}} \frac{\exp \left(-\operatorname{tr}\left(\boldsymbol{\Psi}^{H}\left(\boldsymbol{R}_{\boldsymbol{H}}+\boldsymbol{R}_{\tilde{\boldsymbol{N}}}\right)^{-1} \boldsymbol{\Psi}\right)\right)}{\operatorname{det}^{T}\left(\boldsymbol{R}_{\boldsymbol{H}}+\boldsymbol{R}_{\tilde{\boldsymbol{N}}}\right)} \\
&=\frac{1}{\pi^{T N}} \frac{\exp \!\! \left(\!\!-\!\!\operatorname{tr}\left(\boldsymbol{\Psi}^{H}\!\left(\!\boldsymbol{F}_{\emph{N}}\left(\!\boldsymbol{\Omega}\!+\!\sigma^{2} \boldsymbol{I}_{\emph{N}}\right)\!\! \boldsymbol{F}_{\emph{N}}^{H}\right)^{-1}\!\! \boldsymbol{\Psi}\right)\right)}{\operatorname{det}^{T}\left(\!\boldsymbol{F}_{\emph{N}}\left(\boldsymbol{\Omega}+\sigma^{2} \boldsymbol{I}_{\emph{N}}\right) \boldsymbol{F}_{\emph{N}}^{H}\right)},
\end{aligned}
\end{equation}
where ${\boldsymbol{\Omega}}{\rm{ = diag}}\left\{ {{r_1}, \ldots ,{r_{{N}}}} \right\}$ is the eigenvalue matrix of ${{\boldsymbol{R}}_{\boldsymbol{H}}}$.
According to \eqref{likelihood function}, the likelihood function is only related to the eigenvalues, so we get a new likelihood function:
\begin{equation} \label{likelihood function1}
L(\boldsymbol{\Psi} \mid \boldsymbol{\Lambda})=\frac{1}{\pi^{T N}} \! \frac{\exp \left(\!\!-\operatorname{tr}\!\!\left(\!\boldsymbol{\Psi}^{H}\left(\boldsymbol{F}_{\emph{N}}{\boldsymbol{\Lambda }} \boldsymbol{F}_{\emph{N}}^{H}\right)^{-1} \boldsymbol{\Psi}\right)\right)}{\operatorname{det}^{T}\left(\boldsymbol{F}_{\emph{N}}{\boldsymbol{\Lambda }} \boldsymbol{F}_{\emph{N}}^{H}\right)},
\end{equation}
where
\begin{equation} \label{ML_parameter}
{\boldsymbol{\Lambda }}{\rm{ = }}{\boldsymbol{\Omega }} + {\sigma ^2}{{\boldsymbol{I}}_{\emph{N}}}{\rm{ = }}{\mathop{\rm diag}\nolimits} \left\{ {{\lambda _1}, \ldots ,{\lambda _N}} \right\}.
\end{equation}
Based on \eqref{likelihood function1}, the ML problem is given by
\begin{equation}\label{ML}
\begin{aligned}
{{{\boldsymbol{\Lambda }}^{*}}}&=\underset {\boldsymbol{\Lambda}}{\arg \max } L(\boldsymbol{\Psi} \mid {\boldsymbol{\Lambda }})\\
&=\underset{\boldsymbol{\Lambda}}{\arg \min }\Big[\operatorname{tr}\left(\boldsymbol{F}_{\emph{N}}^{H} \hat{\boldsymbol{R}}_{\tilde{\textbf{Y}}} \boldsymbol{F}_{\emph{N}} \boldsymbol{\Lambda}^{-1}\right)+\sum_{i=1}^{N} \log \lambda_{i}\Big],
\end{aligned}
\end{equation}
where $\boldsymbol{\Lambda}^{*}=\operatorname{diag}\left\{\lambda_{1}^{*}, \ldots, \lambda_{N}^{*}\right\}$ and $\hat{\boldsymbol{R}}_{\tilde{\boldsymbol{Y}}}$ is defined as the sample covariance matrix, which can be expressed as
\begin{equation} \label{R_y}
\hat{\boldsymbol{R}}_{\tilde{\boldsymbol{Y}}}=\frac{1}{T} \boldsymbol{\Psi} \boldsymbol{\Psi}^{H}=\frac{1}{T} \sum_{t=1}^{T} \tilde{\boldsymbol{Y}}_{t} \tilde{\boldsymbol{Y}}_{t}^{H}.
\end{equation}
Further, $\hat{\boldsymbol{R}}_{\tilde{\boldsymbol{Y}}}$ can also be regarded as a circulant matrix. Thus $\hat{\boldsymbol{R}}_{\tilde{\boldsymbol{Y}}}$ can also be diagonalized by DFT matrix as
\begin{equation}\label{Ry_EIG}
{\boldsymbol{\Xi }}{\rm{ = }}{{\boldsymbol{F}}_{\emph{N}}^H}{{{\hat{\boldsymbol{R}} }}_{{{\tilde{\boldsymbol{Y}}}}}}{\boldsymbol{F}}_{\emph{N}} = {\rm{diag\{ }}{\xi _1},...{\rm{,}}{\xi _N}{\rm{\} }}.
\end{equation}
Substituting \eqref{Ry_EIG} into \eqref{ML}, the equivalent optimization problem of \eqref{ML} can be given by
\begin{equation}
\left\{\lambda_{i}^{*}\right\}_{i=1}^{N}=\underset{\lambda_{i}, \forall i=1, \ldots, N}{\arg \min }\left[\sum_{i=1}^{N} \frac{\xi_{i}}{\lambda_{i}}+\log \lambda_{i}\right],
\end{equation}
which is based on the fact that the trace of a matrix is equal to the sum of its eigenvalues.
There is a unique optimal solution $\lambda_{i}^{*} = {\xi _i}$ for each eigenvalue. Therefore, we can intuitively get the optimal solution for \eqref{ML} as
\begin{equation}\label{lambda1}
\boldsymbol{\Lambda}^{*}=\boldsymbol{\Xi}=\operatorname{diag}\left\{\xi_{1}, \ldots, \xi_{N}\right\}.
\end{equation}

Combining \eqref{Ry_EIG} and benefiting from the circulant structure of $ \hat{\boldsymbol{R}}_{\tilde{\boldsymbol{Y}}} $, we can get
\begin{equation}
\lambda_{i}^{*}=\sum_{n=1}^{N}\left[\hat{\boldsymbol{R}}_{\tilde{\boldsymbol{Y}}}\right]_{1, n} \exp \left(\frac{-j 2 \pi n i}{N}\right), i=1,2, \ldots, N,
\end{equation}
which is based on the fact that the eigenvalues of a circulant matrix can be obtained by the DFT of its first row\cite{r10}.

Based on \eqref{ML_parameter} and \eqref{lambda1} , the ML estimator of the eigenvalues for ${{\boldsymbol{R}}_{\boldsymbol{H}}}$ can finally be expressed as
\begin{equation}\label{channel cov eig}
\hat{\boldsymbol{\Omega}}^{\text {ML}}=\boldsymbol{\Lambda}^{*}-\sigma^{2} \boldsymbol{I}_{\emph N}.
\end{equation}
Substituting \eqref{channel cov eig} as the estimate of ${\boldsymbol{\Omega}}$ into \eqref{the LMMSE estimate}, we can get \emph{Theorem 2}.

\subsubsection{Spherical wave properties are not utilized}
We consider a likelihood function similar to \eqref{likelihood function} by
\begin{equation}\label{likelihood function2}
\begin{aligned}
L \left(\boldsymbol{\Psi} \mid \boldsymbol{R} \right) =\frac{1}{\pi^{T N}} \frac{\exp \left(-\operatorname{tr}\left(\boldsymbol{\Psi}^{H} \boldsymbol{R}^{-1} \boldsymbol{\Psi}\right)\right)}{\operatorname{det}^{T}\left(\boldsymbol{R}\right)},
\end{aligned}
\end{equation}
where $\boldsymbol{R}=\boldsymbol{R}_{\boldsymbol{H}}+\boldsymbol{R}_{\tilde{\boldsymbol{N}}}$.
Unfortunately, the eigenvectors of $\boldsymbol{R}$ are unknown because the circulant structure is not utilized. Specifically, $\boldsymbol{R}$ cannot complete the eigenvalue decomposition by the DFT matrix ${{\boldsymbol{F}}\!_{\emph{N}}}$, which leads to the fact that the method in \emph{Theorem 2} is not suitable for this case. We have to use a different method as follows.

Inspired by \cite{r12}, the ML solution of \eqref{likelihood function2} is

\begin{equation}
{\boldsymbol{R}} = \mathop {\arg \max }\limits_{\boldsymbol{R}} L({\boldsymbol{\Psi}}\mid {\boldsymbol{R}}) = \hat{\boldsymbol{R}}_{\tilde{\boldsymbol{Y}}},
\end{equation}
where the definition of $ \hat{\boldsymbol{R}}_{\tilde{\boldsymbol{Y}}} $ is the same as \eqref{R_y}. Therefore, we can obtain the ML estimator of $ \boldsymbol{R}_{\boldsymbol{H}} $ as
\begin{equation}\label{RH_Est}
\hat{\boldsymbol{R}}_{\boldsymbol{H}}^{\text{ML}}=\hat{\boldsymbol{R}}_{\tilde{\boldsymbol{Y}}}-\boldsymbol{R}_{\tilde{\textbf{N}}}.
\end{equation}
Substituting (\ref{RH_Est}) as the estimate of ${\boldsymbol{R}}_{\boldsymbol{H}}$ into (\ref{LMMSE matrix}), we can get the LMMSE estimator as
\begin{equation}
\hat{\boldsymbol{H}}_{t}^{\text{LMMSE}}=\tilde{\boldsymbol{Y}}_{t} \hat{\boldsymbol{R}}_{\tilde{\boldsymbol{Y}}}^{-1} \hat{\boldsymbol{R}}_{\boldsymbol{H}}^{\text{ML}}.
\end{equation}

Naturally, this method will lead to higher complexity because it cannot simplify matrix inversion. The MADs of obtaining $\boldsymbol{A}$ in (\ref{LMMSE matrix}) are $(2 T+3) N^{3}-T N^{2}+N$, but the computational complexity is reduced to $2 T N^{2}+3 N \log _{2} N+(3-T) N+T-1$ in the case where the circulant structure  introduced by SWM is utilized. For example, when $N=256$, the complexity ratio of the former to the latter is almost 293.

The detailed complexity comparison for above four cases is listed in TABLE~\ref{tab1}, which intuitively shows that the proposed methods are suitable for extra-large scale MIMO systems by exploiting the SWP.

\begin{table*}[t]
	\newcommand{\tabincell}[2]{\begin{tabular}{@{}#1@{}}#2\end{tabular}}  
	\caption{Complexity Comparison}
	\begin{center}
		\setlength{\tabcolsep}{1mm}{
			
			\begin{tabular}{c c c c}
				\toprule[0.75pt] 
				Method                           & Additions  & Multiplications & MADs \\ \midrule[0.5pt]
				\rule{0pt}{12pt}	\tabincell{c}{ SWP based LMMSE \\ with known CCM}         & $N + 2N{\log _2}N$ & $N + N{\log _2}N$ & $2N + 3N{\log _2}N$  \\ 	\tabincell{c}{LMMSE with \\ known CCM}      & $\frac{3}{2}{N^3} - \frac{3}{2}{N^2} + N$ &  $\frac{1}{2}{N^3} + \frac{3}{2}{N^2}$ & $3{N^3} + N$  \\ 
				\rule{0pt}{12pt}	\tabincell{c}{SWP based LMMSE \\ with unknown CCM}       & {\makecell[c]{$T N^{2}+2 N \log _{2} N$ \\$+(2-T) N+T-1$}}   & $T N^{2}+N \log _{2} N+N$ & {\makecell[c]{$ 2T{N^2} + 3N{\log _2}N$ \\ $ + (3 - T)N + T - 1  $}}  \\ 
				\rule{0pt}{10pt}	\tabincell{c}{LMMSE with \\ unknown CCM}   &{\makecell[c]{$(T + \frac{3}{2}){N^3}$ \\ $ - (T + \frac{3}{2}){N^2} + N$}}   & $ (T + \frac{3}{2}){N^3} + \frac{3}{2}{N^2} $ & $ (2T + 3){N^3} - T{N^2} + N $  \\ 
				\bottomrule[0.75pt]
		\end{tabular}}
		\label{tab1}
	\end{center}
\end{table*}

\begin{figure*}[tbp]
	\centering
	\includegraphics[width=0.8\textwidth]{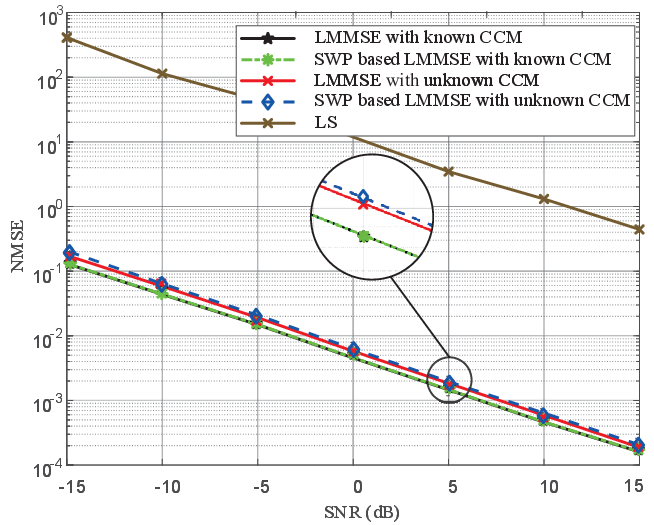}\\
	\caption{NMSE comparison of different estimators as a function of SNR}
	\label{NMSE}
\end{figure*}

\section{Simulation Results}
In this section, we present the numerical results to evaluate the performance of the proposed algorithm. We assume that the noise power ${{\sigma }^{2}}$ is known, and set the carrier frequency $f=\text{100GHz}$, ${R_{t}=R_{r}=0.5 \mathrm{m}}$, $d=100\mathrm{m}$, $T=10$. 

In Fig.~\ref{NMSE}, we compare the normalized mean-square error (NMSE) of LMMSE in four cases on the condition of whether the circulant structure  introduced by the SWM is utilized and whether the CCM is known. The number of antennas $N=512$ is considered. As a comparison, we also show the NMSE for least squares (LS) estimator. As expected, we can see that the gap between the LS estimator and other estimators. The results show that the low-complex SWP based LMMSE with known CCM achieves the same performance as the LMMSE without using SWP. In the case of unknown CCM, due to the estimation error, there is a slight deviation from the case with the known CCM, but the SWP based LMMSE estimator still has a significant advantage because of its  low complexity with minimal performance loss.

\begin{figure*}[tbp]
	\centering
	\includegraphics[width=0.8\textwidth]{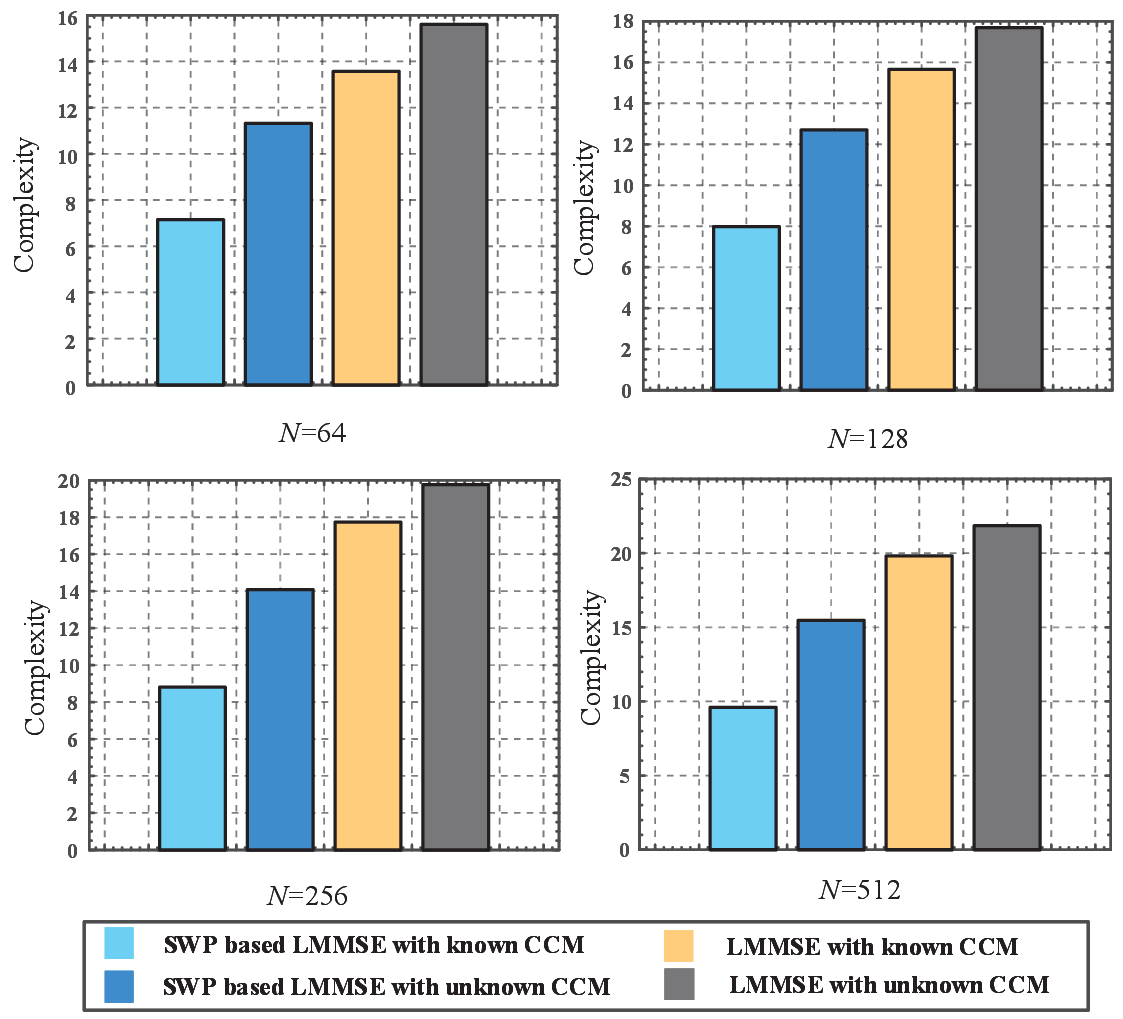}\\
	\caption{Computational complexity versus number of antennas}
	\label{MADs}
\end{figure*}

The comparison of complexity is shown in Fig.~\ref{MADs}, which shows the logarithmic computational complexity (log(MADs)) in the same four cases as  Fig.~\ref{NMSE}. Compared with the case that the circulant structure is not utilized, the computational complexity will be greatly reduced by using the circulant structure introduced by the SWM.

\section{Conclusion}
We presented a low-complex LMMSE channel estimation for the extra-large scale MIMO based on the SWP. By using the circulant properties of the channel matrix  introduced by the SWM, the computation complexity of the LMMSE channel
estimation with known and unknown CCM is greatly reduced. The proposed channel estimator is suitable for the extra-large scale MIMO systems in the THz band.










\printcredits

\section*{Declaration of competing interest}
The authors declare that they have no known competing financial interests or personal relationships that could have appeared to influence the work reported in this paper.

\section*{Acknowledgement}
This work is supported by the Science and Technology Research Program of Chongqing Municipal Education Commission KJQN202100649, China Postdoctoral Science Foundation 2019M651649, Jiangsu Planned Projects for Postdoctoral Research Funds 2018K041c, National Natural Science Foundation of China 61701062.

\bibliographystyle{elsarticle-num}

\bibliography{bibsample}

\begin{thebibliography}{10}
\expandafter\ifx\csname url\endcsname\relax
  \def\url#1{\texttt{#1}}\fi
\expandafter\ifx\csname urlprefix\endcsname\relax\def\urlprefix{URL }\fi
\expandafter\ifx\csname href\endcsname\relax
  \def\href#1#2{#2} \def\path#1{#1}\fi

\bibitem{Noh}
N.~Tanner, J.~Wait, C.~Farrar, H.~Sohn, Channel estimation techniques for
  {RIS}-assisted communication: Millimeter-wave and sub-{TH}z systems, IEEE
  Veh. Technol. Mag. 17~(2) (2022) 64--73,
  \url{http://dx.doi.org/10.1109/MVT.2022.3158765}.

\bibitem{pu}
X.~Pu, S.~Shao, Y.~Tang, Optimal 2 $\times$ 2 antenna placement for short-range
  communications, IEEE Commun. Lett. 17~(8) (2013) 1560--1563,
  \url{http://dx.doi.org/10.1109/LCOMM.2013.070113.130758}.

\bibitem{Wu}
S.~Wu, C.~X. Wang, e.~H.~M.~Aggoune, M.~M. Alwakeel, X.~You, A general 3-{D}
  non-stationary {5G} wireless channel model, IEEE Trans. Commun. 66~(7) (2018)
  3065--3078, \url{http://dx.doi.org/10.1109/TCOMM.2017.2779128}.

\bibitem{Pu8}
X.~Pu, Q.~Chen, S.~Shao, R.~Chai, Y.~Tang, Transmit design for short-range
  {MIMO} channels with the spherical-wave model, IEEE Commun. Lett. 21~(8)
  (2017) 1875--1878, \url{http://dx.doi.org/10.1109/LCOMM.2017.2695604}.

\bibitem{Ai}
X.~Yin, S.~Wang, N.~Zhang, B.~Ai, Scatterer localization using large-scale
  antenna arrays based on a spherical wave-front parametric model, IEEE Trans.
  Wireless Commun. 16~(10) (2017) 6543--6556,
  \url{http://dx.doi.org/10.1109/TWC.2017.2725260}.

\bibitem{Magoarou}
L.~L. Magoarou, A.~L. Calvez, S.~Paquelet, Massive {MIMO} channel estimation
  taking into account spherical waves, in: IEEE 20th International Workshop on
  Signal Processing Advances in Wireless Communications (SPAWC), Cannes,
  France, 2019, pp. 1--5, \url{http://dx.doi.org/10.1109/SPAWC.2019.8815472}.

\bibitem{Han7}
Y.~Han, S.~Jin, C.~K. Wen, X.~Ma, Channel estimation for extremely large-scale
  massive {MIMO} systems, IEEE Wireless Commun. Lett. 9~(5) (2020) 633--637,
  \url{http://dx.doi.org/10.1109/LWC.2019.2963877}.

\bibitem{Savaus}
V.~Savaus, Y.~Louët, {LMMSE} channel estimation in {OFDM} context: A review,
  IET Signal Process. 11~(2) (2017) 123--134,
  \url{http://dx.doi.org/10.1049/iet-spr.2016.0185}.

\bibitem{Jacobs2009}
L.~Jacobs, M.~Moeneclaey, Effect of {MMSE} channel estimation on ber
  performance of orthogonal space-time block codes in {R}ayleigh fading
  channels, IEEE Trans. Commun. 57~(5) (2009) 1242--1245,
  \url{http://dx.doi.org/10.1109/TCOMM.2009.05.070455}.

\bibitem{Taesang2004}
T.~Yoo, A.~Goldsmith, Capacity of fading {MIMO} channels with channel
  estimation error, in: IEEE 20th International Workshop on Signal Processing
  Advances in Wireless Communications (SPAWC), Vol.~2, Paris, France, 2004, pp.
  808--813, \url{http://dx.doi.org/10.1109/ICC.2004.1312613}.

\bibitem{wu2021}
H.~Wu, {LMMSE} channel estimation in {OFDM} systems: A vector quantization
  approach, IEEE Commun. Lett. 25~(6) (2021) 1994--1998,
  \url{http://dx.doi.org/10.1109/LCOMM.2021.3059776}.

\bibitem{Wan}
Q.~Wan, J.~Fang, H.~Duan, Z.~Chen, H.~Li, Generalized bussgang {LMMSE} channel
  estimation for one-bit massive {MIMO} systems, IEEE Trans. Wireless Commun.
  19~(6) (2020) 4234--4246, \url{http://dx.doi.org/10.1109/TWC.2020.2981599}.

\bibitem{Cheng}
X.~Cheng, B.~Xia, K.~Xu, S.~Li, Bayesian channel estimation and data detection
  in oversampled {OFDM} receiver with low-resolution {ADC}, IEEE Trans.
  Wireless Commun. 20~(9) (2021) 5558--5571,
  \url{http://dx.doi.org/10.1109/TWC.2021.3068484}.

\bibitem{Edfors}
O.~Edfors, M.~Sandell, J.~J. van~de Beek, S.~K. Wilson, P.~O. Borjesson, {OFDM}
  channel estimation by singular value decomposition, IEEE Trans. Commun.
  46~(7) (1998) 931--939, \url{http://dx.doi.org/10.1109/26.701321}.

\bibitem{Zhou}
W.~Zhou, W.~Lam, A fast {LMMSE} channel estimation method for {OFDM} systems,
  EURASIP J. Wirel. Commun. Netw. 2009 (2009) 1--13,
  \url{http://dx.doi.org/10.1155/2009/752895}.

\bibitem{r11}
D.~Neumann, T.~Wiese, W.~Utschick, Learning the {MMSE} channel estimator, IEEE
  Trans. Signal Process. 66~(11) (2018) 2905--2917,
  \url{http://dx.doi.org/10.1109/TSP.2018.2799164}.

\bibitem{JSong}
H.~J. Song, N.~Lee, Terahertz communications: Challenges in the next decade,
  IEEE Trans. Terahertz Sci. Technol. 12~(2) (2022) 105--117,
  \url{http://dx.doi.org/10.1109/TTHZ.2021.3128677}.

\bibitem{Do}
H.~Do, S.~Cho, J.~Park, H.~J. Song, N.~Lee, A.~Lozano, Terahertz line-of-sight
  {MIMO} communication: Theory and practical challenges, IEEE Commun. Mag.
  59~(3) (2021) 104--109, \url{http://dx.doi.org/10.1109/MCOM.001.2000714}.

\bibitem{PWang}
P.~Wang, Y.~Li, B.~Vucetic, Millimeter wave communications with symmetric
  uniform circular antenna arrays, IEEE Commun. Lett. 18~(8) (2014) 1307--1310,
  \url{http://dx.doi.org/10.1109/LCOMM.2014.2332334}.

\bibitem{Jing}
H.~Jing, W.~Cheng, W.~Zhang, H.~Zhang, Optimal {UCA} design for {OAM} based
  wireless backhaul transmission, in: IEEE International Conference on
  Communications (ICC), Dublin, Ireland, 2020, pp. 1--6,
  \url{http://dx.doi.org/10.1109/ICC40277.2020.9148901}.

\bibitem{Long}
W.~X. Long, R.~Chen, M.~Moretti, J.~Xiong, J.~Li, Joint spatial division and
  coaxial multiplexing for downlink multi-user {OAM} wireless backhaul, IEEE
  Trans. Broadcast. 67~(4) (2021) 879--893,
  \url{http://dx.doi.org/10.1109/TBC.2021.3081869}.

\bibitem{Torkildson}
E.~Torkildson, U.~Madhow, M.~Rodwell, Indoor millimeter wave {MIMO}:
  Feasibility and performance, IEEE Trans. Wireless Commun. 10~(12) (2011)
  4150--4160, \url{http://dx.doi.org/10.1109/TWC.2011.092911.101843}.

\bibitem{M}
M.~Biguesh, A.~B. Gershman, Training-based {MIMO} channel estimation: A study
  of estimator tradeoffs and optimal training signals, IEEE Trans. Signal
  Process. 54~(3) (2006) 884--893,
  \url{http://dx.doi.org/10.1109/TSP.2005.863008}.

\bibitem{r10}
P.~Davis, Circulant matrices, John Wiley, New York, 1979.

\bibitem{r12}
A.~Dembo, The relation between maximum likelihood estimation of structured
  covariance matrices and periodograms, IEEE Trans. Acoust., Speech, Signal
  Process. 34~(6) (1986) 2905--2917,
  \url{http://dx.doi.org/10.1109/TASSP.1986.1164969}.

\end{thebibliography}

\bio{}
\endbio

\end{document}